\documentclass{bmcart}

\RequirePackage{natbib}
\RequirePackage{hyperref}
\usepackage[utf8]{inputenc} %unicode support
\usepackage{geometry}
\usepackage{tikz}
\usetikzlibrary{shapes,arrows,positioning}

\startlocaldefs
\usepackage{graphicx}
\usepackage{multicol}
\graphicspath{{C:/Users/Svenja/Desktop/bmc_template/rausgeschickt/plots/}}
\usepackage{makecell}
\usepackage{booktabs}
\usepackage{verbatim}
\endlocaldefs

\providecommand{\new}[1]{{#1}}

\begin{document}

%%% Start of article front matter
\begin{frontmatter}

\begin{fmbox}
\dochead{Research}

%%%%%%%%%%%%%%%%%%%%%%%%%%%%%%%%%%%%%%%%%%%%%%
%%                                          %%
%% Enter the title of your article here     %%
%%                                          %%
%%%%%%%%%%%%%%%%%%%%%%%%%%%%%%%%%%%%%%%%%%%%%%

\title{Likelihood-based random-effects meta-analysis with few studies: Empirical and simulation studies}

%%%%%%%%%%%%%%%%%%%%%%%%%%%%%%%%%%%%%%%%%%%%%%
%%                                          %%
%% Enter the authors here                   %%
%%                                          %%
%% Specify information, if available,       %%
%% in the form:                             %%
%%   <key>={<id1>,<id2>}                    %%
%%   <key>=                                 %%
%% Comment or delete the keys which are     %%
%% not used. Repeat \author command as much %%
%% as required.                             %%
%%                                          %%
%%%%%%%%%%%%%%%%%%%%%%%%%%%%%%%%%%%%%%%%%%%%%%

\author[
   addressref={AMS,HBG},                   % id's of addresses, e.g. {aff1,aff2}
   noteref={n1},                        % id's of article notes, if any
   email={seide@imbi.uni-heidelberg.de}   % email address
]{\inits{SE}\fnm{Svenja E} \snm{Seide}}
\author[
   addressref={AMS},
   %noteref={n2},                        % id's of article notes, if any
   email={christian.roever@med.uni-goettingen.de},
   corref={AMS}        % id of corresponding address, if any
]{\inits{C}\fnm{Christian} \snm{R\"over}}
\author[
   addressref={AMS},
   noteref={n3},                        % id's of article notes, if any
   email={tim.friede@med.uni-goettingen.de}
]{\inits{T}\fnm{Tim} \snm{Friede}}

%%%%%%%%%%%%%%%%%%%%%%%%%%%%%%%%%%%%%%%%%%%%%%
%%                                          %%
%% Enter the authors' addresses here        %%
%%                                          %%
%% Repeat \address commands as much as      %%
%% required.                                %%
%%                                          %%
%%%%%%%%%%%%%%%%%%%%%%%%%%%%%%%%%%%%%%%%%%%%%%

\address[id=AMS]{%                                 % unique id
  \orgname{Department of Medical Statistics, University Medical Center G\"ottingen}, % university, etc
  \street{Humboldtallee 32},                       %
  \postcode{37073},                                % post or zip code
  \city{G\"ottingen},                              % city
  \cny{Germany},                                   % country
}
\address[id=HBG]{%                                 % unique id
  \orgname{Institute of Medical Biometry and Informatics, Heidelberg University Hospital}, % university, etc
  \street{Im Neuenheimer Feld 130.3},              %
  \postcode{69120},                                % post or zip code
  \city{Heidelberg},                               % city
  \cny{Germany},                                   % country
}

%%%%%%%%%%%%%%%%%%%%%%%%%%%%%%%%%%%%%%%%%%%%%%
%%                                          %%
%% Enter short notes here                   %%
%%                                          %%
%% Short notes will be after addresses      %%
%% on first page.                           %%
%%                                          %%
%%%%%%%%%%%%%%%%%%%%%%%%%%%%%%%%%%%%%%%%%%%%%%

\begin{artnotes}
%\note{Sample of title note}     % note to the article
\note[id=n1]{\texttt{seide@imbi.uni-heidelberg.de} } % note, connected to 
 \note[id=n3]{\texttt{tim.friede@med.uni-goettingen.de}} % note, connected to author
\end{artnotes}

%\end{fmbox}% comment this for two column layout

%%%%%%%%%%%%%%%%%%%%%%%%%%%%%%%%%%%%%%%%%%%%%%
%%                                          %%
%% The Abstract begins here                 %%
%%                                          %%
%% Please refer to the Instructions for     %%
%% authors on http://www.biomedcentral.com  %%
%% and include the section headings         %%
%% accordingly for your article type.       %%
%%                                          %%
%%%%%%%%%%%%%%%%%%%%%%%%%%%%%%%%%%%%%%%%%%%%%%

%\rightline{\Large \color{red} \hspace{1ex} \texttt{+++ DRAFT {\today} +++}}
\begin{abstractbox}

\begin{abstract} % abstract
  \parttitle{Background} 
  Standard random-effects meta-analysis methods perform poorly when applied to
  few studies only. Such settings however are commonly encountered in
  practice. It is unclear, whether or to what extent
  small-sample-size behaviour can be improved by more sophisticated
  modeling.

  \parttitle{Methods} We consider likelihood-based methods, the
  DerSimonian-Laird approach, Empirical Bayes, several adjustment
  methods and a fully Bayesian approach. Confidence intervals are
  based on a normal approximation, or on adjustments based on the
  Student-$t$-distribution. In addition, a linear mixed model and two
  generalized linear mixed models \new{(GLMMs)} assuming binomial or Poisson
  distributed numbers of events per study arm are considered for
  pairwise binary meta-analyses.  We extract an empirical data set of
  40 meta-analyses from recent reviews published by the German
  Institute for Quality and Efficiency in Health Care (IQWiG).
  Methods are then compared empirically and as well as in a simulation
  study, based on few studies, imbalanced study sizes, and considering
  odds-ratio (OR) and risk ratio (RR) effect sizes.  
% Type-I errors,
  \new{Coverage probabilities}
  and interval widths 
% and bias 
  for the combined effect estimate are
  evaluated to compare the different approaches.

  \parttitle{Results} Empirically, a majority of the identified
  meta-analyses include only 2 studies. Variation of methods or effect
  measures affects the estimation results. 
  In the simulation study, coverage probability is, in the presence of
  heterogeneity and few studies, mostly below the nominal level for all
  frequentist methods based on normal approximation, in particular
  when sizes in meta-analyses are not balanced, but improve when
  confidence intervals are adjusted.
  Bayesian methods result in better coverage than the
  frequentist methods with normal approximation in all
  scenarios\new{, except for some cases of very large heterogeneity where the coverage is slightly lower}. Credible intervals are empirically and in the simulation
  study wider than unadjusted confidence intervals, but considerably
  narrower than adjusted ones\new{, with some exceptions when considering RRs and small numbers of patients per trial-arm}. Confidence intervals based on the
  \new{GLMMs} are, in general, slightly narrower
  than those from other frequentist methods. 
%  Certain 
  \new{Some}
  methods turned
  out impractical due to frequent numerical problems.

  \parttitle{Conclusions} In the presence of between-study
  heterogeneity, especially with unbalanced study sizes, caution is
  needed in applying meta-analytical methods to few studies, as either
  coverage probabilities might be compromised, or intervals are
  inconclusively wide. Bayesian estimation with a sensibly chosen
  prior for between-trial heterogeneity may offer a promising
  compromise.
\end{abstract}

%%%%%%%%%%%%%%%%%%%%%%%%%%%%%%%%%%%%%%%%%%%%%%
%%                                          %%
%% The keywords begin here                  %%
%%                                          %%
%% Put each keyword in separate \kwd{}.     %%
%%                                          %%
%%%%%%%%%%%%%%%%%%%%%%%%%%%%%%%%%%%%%%%%%%%%%%


% MSC classifications codes, if any
%\begin{keyword}[class=AMS]
%\kwd[Primary ]{}
%\kwd{}
%\kwd[; secondary ]{}
%\end{keyword}
\end{abstractbox}
\end{fmbox}% uncomment this for twcolumn layout

\end{frontmatter}

%%%%%%%%%%%%%%%%%%%%%%%%%%%%%%%%%%%%%%%%%%%%%%
%%                                          %%
%% The Main Body begins here                %%
%%                                          %%
%% Please refer to the instructions for     %%
%% authors on:                              %%
%% http://www.biomedcentral.com/info/authors%%
%% and include the section headings         %%
%% accordingly for your article type.       %%
%%                                          %%
%% See the Results and Discussion section   %%
%% for details on how to create sub-sections%%
%%                                          %%
%% use \cite{...} to cite references        %%
%%  \cite{koon} and                         %%
%%  \cite{oreg,khar,zvai,xjon,schn,pond}    %%
%%  \nocite{smith,marg,hunn,advi,koha,mouse}%%
%%                                          %%
%%%%%%%%%%%%%%%%%%%%%%%%%%%%%%%%%%%%%%%%%%%%%%

%%%%%%%%%%%%%%%%%%%%%%%%% start of article main body
% <put your article body there>

%%%%%%%%%%%%%%%%
%% Background %%
%%%%%%%%%%%%%%%%

\section*{Background}
%%%
%%%  For author guidelines, see also: 
%%%  https://bmcmedresmethodol.biomedcentral.com/submission-guidelines/preparing-your-manuscript/research-article
%%%
  Meta-analyses of few 
% and very few 
  studies are common in
  practice. For instance, a review of the Cochrane Library revealed
  that half of the meta-analyses reported in the Cochrane Library are
  conducted with two or three studies \citep{TurnerEtAl2012}. However,
  standard random-effects meta-analysis methods perform poorly when
  applied to 
% (very) 
  few studies only
  \citep{RoeverKnappFriede2015,FriedeRoeverWandelNeuenschwander2017a}. It
  is unclear, whether or to what extent small-sample-size behaviour
  can be improved by more sophisticated modeling. Bayesian
  random-effects meta-analyses with weakly informative priors for the
  between-study heterogeneity have been proposed for this
  setting \citep{HigginsThompsonSpiegelhalter2009} and their
  performance has been found to be satisfactory in numerical
  applications and simulations
  \citep{FriedeRoeverWandelNeuenschwander2017a,
    FriedeRoeverWandelNeuenschwander2017b}. Other alternative
  approaches including likelihood based methods have been mentioned as
  potential remedies \citep{BenderEtAl2018}.

  In meta-analyses commonly a two-stage approach is applied. In the
  first step, data from the individual studies are analyzed resulting
  in effect estimates with standard errors. These are then combined in
  a second step. As individual patient data (IPD) are not generally
  available and effects with standard errors can typically extracted
  from publications, this two-stage approach makes a lot of sense from
  a practical point of view. With binary data, however, the individual
  patient data are summarized by $2\!\times\!2$ frequency tables and are
  usually readily available from publications
  \citep{bohning2008meta}. Therefore, preference might be given to
  one-stage approaches in this setting over the commonly applied
  two-stage approach.
%  as the one-stage approach is generally more flexible and more exact \citep{DebrayEtAl2013}. 
  However, numerical
  differences between the one-stage and two-stage approaches have been
  found to be small in a simple Gaussian model
  \citep{MorrisEtAl2018}. If differences are observed, these arise
  mostly for differing models
  \citep{MathewNordstroem2010,BurkeEnsorRiley2017} or relate not to
  the main effects but interactions \citep{Kontopantelis2018}.
  So, while a simpler two-stage model is often sufficient (especially
  in case of many studies and non-rare events), a one-stage model may
  on the other hand be expected to be more flexible and more exact
  \citep{DebrayEtAl2013}.  A Bayesian approach may be more suitable
  especially in cases of few studies
  \citep{HigginsThompsonSpiegelhalter2009,FriedeRoeverWandelNeuenschwander2017a,FriedeRoeverWandelNeuenschwander2017b}.
  For a more detailed discussion of common
  models for binary data, see also Jackson \textit{et al.}
  \citep{JacksonEtAl2018}.

  Although some model and method comparison studies appeared recently
  \citep{InthoutIoannidisBorm2014,JacksonEtAl2018}, a systematic
  evaluation and comparison of the various methods is lacking in the
  context of 
% (very) 
  few studies. Here we intend to close this gap by
  an empirical study and comprehensive simulations.

  This manuscript is structured as follows. In the following section
  we summarize the meta-analysis approaches compared, the extraction
  of the empirical data set and the setup of the simulation
  study. Then the results of the empirical study as well as of the
  simulation study are presented. We close with a brief discussion and
  some conclusions.

%%%%%%%%%%%%%%%%
%% Methods    %%
%%%%%%%%%%%%%%%%
\section*{Methods}
\subsection*{Modeling approaches}
  In the following, we will consider meta-analyses based on binary
  endpoints, where each study's outcome may be summarized in a
  \mbox{$2\!\times\!2$ table} giving the the numbers of participants
  with and without an event in both study arms.

\subsection*{Normal-normal hierarchical model (NNHM)}
  \subsubsection*{Model specification}
  Traditionally, meta-analytical methods often follow a contrast-based
  summary measure approach which is based on the log-transformed
  empirical estimates of the outcome measure and their standard
  errors, and assuming an approximate normal likelihood
  \citep{Fleiss1993}.

  In a common situation in random-effects meta-analysis,
  $k$~independent studies are available in which the treatment
  effect~$\theta_i$ is the parameter of interest ($i = 1,2, \ldots,
  k$). From each study, an \new{effect} estimate~$\hat{\theta}_i$ with its
  \new{estimated} variance \new{(squared standard error)}~$\sigma_i^2$ is provided for this treatment effect. It is
  then assumed that $\hat{\theta}_i$~follows a normal distribution
  centered around the unknown true treatment effect~$\theta_i$, with
  the variance~$\sigma_i^2$ accounting for the measurement
  uncertainty, or within-study variation.
  \new{Although $\sigma_i^2$ usually only is an estimate, it is commonly treated as known.}
  The~$\theta_i$ may vary
  across study populations around a global mean~$\mu$ due to the
  between-study heterogeneity~$\tau$. After integrating out the
  parameters~$\theta_i$, the marginal model can be expressed as
  \begin{equation} \label{eqn:nnhm}
    \hat{\theta}_i \;\sim\; \mathcal{N}(\mu, \,\sigma_i^2 + \tau^2) \mbox{.}
  \end{equation}
  This model is commonly applied to both log-transformed risk ratio
  (RR) or odds ratio (OR) measures of treatment effect for binary data
  $\hat{\theta}_i$ \citep{HedgesOlkin,HartungKnappSinha}; it is
  denoted as ``model~1'' in the investigation by Jackson \textit{et al.}
  \citep{JacksonEtAl2018}.

  \subsubsection*{Inference}
  We will consider frequentist and Bayesian approaches to inference
  within the generic NNHM\@.  In the frequentist approaches, an
  estimate of the between-study heterogeneity~$\tau$ is usually
  required first. Different estimators are available; in the following
  we consider the commonly used DerSimonian-Laird (DL)
  \citep{derSimonian1986}, maximum likelihood (ML), restricted maximum
  likelihood (REML) \citep{viechtbauer2005, raudenbush2009} and
  empirical Bayes (EB) estimators, the latter also being known as the
  Paule-Mandel estimator \citep{morris1996, paule1982}. Based on an
  estimate of this heterogeneity~$\hat{\tau}$, the mean effect
  estimates are determined in a subsequent step by conditioning on
  the~$\hat{\tau}$ value as if it were known.

  The fully Bayesian estimation within the NNHM framework is done
  using three different prior specifications for the between-study
  heterogeneity~($\tau$). Uncertainty in the heterogeneity is
  naturally accounted for when estimating the combined treatment
  effect~$\mu$ by marginalisation. Especially if the number of studies
  is small, however, the choice of priors matters, as has been
  discussed by Turner \textit{et al.} \citep{TurnerEtAl2015}, Dias
  \textit{et al.}  \citep[][Sec.~6.2]{dias2014}, or R\"{o}ver
  \citep{Roever2017}. We follow Friede \textit{et al.}
  \citep{FriedeRoeverWandelNeuenschwander2017a} and Spiegelhalter
  \textit{et al.}  \citep[][Sec.~5.7]{spiegelhalter2004} and consider
  two half-normal priors with scales~$0.5$ and~$1.0$ 
%  and a uniform (\mbox{$0,\,4$}) prior 
  for the between-study heterogeneity.  
  These specifications include up to ``\emph{fairly high}'' and
  ``\emph{fairly extreme}'' heterogeneity
  \citep[Sec.~5.7.3]{spiegelhalter2004}, and they also span the range
  of values considered in the simulations (see
  Table~\ref{tab:Heterogeneity}).
  In all of these approaches risk ratios (RR) and odds ratios (OR) can
  be used as the treatment effect.

\subsection*{Generalized linear mixed models (GLMM)}
  \subsubsection*{Models}
  The statistical model may also be based directly on the count data,
  using either a binomial or a Poisson assumption on the numbers of
  events per study arm. Generalized linear mixed models \new{(GLMMs)} may then be
  fitted to the data, using a logarithmic link for Poisson rates or a
  logit link for proportions. Treatment effects may be modeled based
  on ORs or RRs, and random effects may be included at several stages
  in order to account for heterogeneity. In addition, we also consider
  some approximate variants of these models. The models used are
  outlined briefly below; most of these are also discussed in more
  detail by Jackson \textit{et al.} \citep{JacksonEtAl2018}.

  \subsubsection*{Model specification and inference}
  If a Poisson distribution is assumed for the
  number of events per arm and study, a log-link will be used to model
  the RR\@. Following B\"ohning \textit{et al.}
  \citep[][Ch. 2]{bohning2008meta} this model is estimated using the
  profile likelihood; in the following, this model will be denoted as
  the ``PN-PL'' model.

  For binomially distributed numbers of events per study arm, a
  logit-link will be applied to model ORs in a logistic
  regression. Four different specifications are included in the
  comparison. Unconditional logistic regression with fixed and random
  study-specific nuisance parameters as discussed by Turner
  \textit{et al.} \citep{turner2000} are considered (``UM.FS'' and
  ``UM.RS'', respectively, in the following). These correspond to
  models~4 and~5 in Jackson
  \textit{et al.} \citep{JacksonEtAl2018}.

  In addition, we follow van~Houwelingen \textit{et al.}
  \citep{vanhouwelingen1993} in using a conditional logistic approach,
  where the total number of events per study is conditioned upon, in
  order to avoid the need to also model their variability
  \citep{viechtbauer2005}. The likelihood of this conditional model
  can be described using Fisher's non-central hypergeometric
  distribution \citep{vanhouwelingen1993} (``CM.EL'' in the following,
  and corresponding to model~7 in \citep{JacksonEtAl2018}).

  Fisher's non-central hypergeometric distribution may be approximated
  by a binomial distribution, if the number of cases is small compared
  to the overall participants in that study \citep{stijnen2010}; this
  model specification will be denoted by ``CM.AL'' in the following
  (approximate version of model~7 in
  \citep[Sec.~3.7.2]{JacksonEtAl2018}).  All of the logistic
  regression models are fitted using maximum likelihood.

\subsection*{Confidence and credible intervals combined effects}
  The 95\% credible intervals in the Bayesian estimation and
  confidence intervals in the frequentist approaches are estimated for
  the combined treatment effect~$\mu$. The narrowest 95\% highest
  posterior density intervals are used in the Bayesian estimation. For
  the construction of confidence intervals, Wald-type intervals based
  on normal quantiles are considered, which are known to be
  anti-conservative when the number of studies is small or
  non-negligible amounts of heterogeneity are present
  \citep{FriedeRoeverWandelNeuenschwander2017a, RoeverKnappFriede2015,
    hartung2001, hartung2001b}. To account for this behaviour,
  confidence intervals are in addition constructed using Student's
  t-distribution in case of the GLMMs, and the
  Hartung-Knapp-Sidik-Jonkman (HKSJ) adjustment \citep{hartung2001,
    hartung2001b, sidik2002} in case of the NNHM\@. The HKSJ-adjusted
  intervals tend to be wider than the Wald-type intervals, although
  this is not strictly the case
  \citep{RoeverKnappFriede2015,sidik2002,knapp2003}. Knapp and Hartung
  \citep{knapp2003} proposed a modification of the
  Hartung-Knapp-Sidik-Jonkman adjustment (mHKSJ) correcting
  HKSJ-adjusted intervals in the cases where they are
  counterintuitively narrow. These modified confidence intervals
  are also considered.

\subsection*{$I^2$ as measure of between-study heterogeneity}
  The ``relative amount of between-study heterogeneity'' can be
  expressed in terms of the measure~$I^2$, which expresses the the
  between-study variance ($\tau^2$) in relation to the overall
  variance ($\tilde{\sigma}^2$) \citep{higgins2002}, which is stated
  as
  \begin{equation}
    I^2 = \frac{\hat{\tau}^2}{\tilde{\sigma}^2 + \hat{\tau}^2} \mbox{.}
  \end{equation}
  In the calculation of $I^2$, a ``typical'' $\tilde{\sigma}^2$ value
  is required as an estimate of the within-study variances
  $\sigma_i^2$. Higgins and Thompson \citep{higgins2002} suggest a
  weighted average of the individual within-study variances as
  ``typical'' value. This, together with the fact that the $I^2$ is
  bounded between zero and one, permits the interpretation of
  heterogeneity magnitude as a relative percentage. The $I^2$ is used
  to set the amount of heterogeneity in the simulation study. Hoaglin
  \citep{hoaglin2016} remarks that the probability for observing a
  moderate (estimated) $I^2$ even in the absence of heterogeneity is
  dependent on the number of studies included and is not
  negligible. As the $I^2$ expresses the between-study variation
  relative to the total variation, the same values of~$\tau$ may
  lead to different values of~$I^2$, depending on the precision of the
  underlying studies and should therefore always be interpreted as a
  relative measure \citep{borenstein2017}.

\subsection*{Extraction of the empirical data set}
  A data set of 40~meta-analyses was extracted from publications of
  the German Institute for Quality and Efficiency in Health Care
  (IQWiG)\@. IQWiG publications were searched chronologically for
  meta-analyses of binary data in April~2017 starting with the most
  recent available ones and reaching back to March~2012. In total,
  521~documents were screened, including all document types in the
  search. If a detailed and a short version of a document existed,
  only the detailed version was considered. From documents including
  at least one meta-analysis of binary data, the first one was
  extracted to obtain a realistic data set with respect to the number
  of studies typically included in a meta-analysis and the sample
  sizes of those studies. Meta-analyses involving studies with zero
  events in one or more arms were excluded from the data set for
  better comparability of the evaluated methods.

\subsection*{Simulation procedure}
  To compare properties of the investigated approaches to
  meta-analysis, we conducted a Monte-Carlo simulation adapting the
  setup from IntHout \textit{et al.} \citep{InthoutIoannidisBorm2014}
  who described the simulation of \mbox{$2\!\times\!2$ tables}. In
  deviation from IntHout \textit{et al.}
  \citep{InthoutIoannidisBorm2014}, series of trials with up to 10
  studies were simulated, and each series was repeated only
  2000~times. Three different designs were considered, where in the
  first one all studies were of equal size, one study was ten times
  larger than the other studies in the second, and one study was only
  a tenth of the size of the other studies in the third design. It
  should be noted however that this ratio corresponds to extreme, but
  not unrealistic cases, as is also illustrated in the right panel of
  Figure~\ref{fig:descriptiveIQWiG}.  \new{The (less common) case
  of equal sizes is of interest here, as this is where we expect the
  HKSJ methods to perform best
  \citep{InthoutIoannidisBorm2014,RoeverKnappFriede2015}}.

  To generate dichotomous outcomes, $p_0$ and $I^2$ have to be set in
  advance. Considered values of the $I^2$ correspond to levels of no,
  low, moderate, high and very high heterogeneity, respectively
  \citep{HigginsEtAl2003}. Note however, that the same $I^2$ value may
  correspond to different values of between-study heterogeneity~$\tau$ 
  depending on the effect measure used, and on whether or not
  study sizes are balanced; the resulting $\tau$~values are shown in
  Table~\ref{tab:Heterogeneity}.  From the $\tau$~values one can see
  that in some of the scenarios, the $I^2$~settings imply
  unrealistically large absolute heterogeneity
  \citep[Sec.~5.7.3]{spiegelhalter2004}, which needs to be considered
  in the interpretation.  This would be true for instance for odds
  ratios with $I^2$ in the range of~0.75 and~0.90 and one
  \new{small} study
  when $\tau$~is roughly in the range of~1 to~2 (see
  Table~\ref{tab:Heterogeneity}).  The baseline event rate ($p_0$)
  needs to be set as an additional parameter and varies from~0.1
  to~0.9 in steps of~0.2. The treatment effect $\theta_i$ is set to
  unity for both RR and~OR, which corresponds to the absence of an
  effect. 
  \new{Note that while for meta-analyses of continuous (or, more specifically, normally distributed) endpoints the magnitude of the simulated treatment effect ($\theta_i$) should not affect performance, e.g.\ for binomial counts it may make a difference, as it affects the chances of observing few or zero events in the treatment arm. However, since we chose not to focus on rare-event issues, and in order to keep the number of simulation scenarios manageable, only the case of \emph{no effect} was investigated.}
  For every combination of the simulation parameters 2000
  repetitions are simulated.
  In case zero event counts occurred, for the models based on the
  NNHM, a continuity correction of~0.5 was added to all cells of the
  affected study's contingency table. Zero counts, however, were rare
  in the scenarios considered.
  The simulation scenarios are also summarized in
  Table~\ref{tab:SimulationParameter}.
  \new{For more details on the simulation procedure see also IntHout \textit{et al.} \citep{InthoutIoannidisBorm2014} and Table~\ref{tab:simulationprocedure} below.}
  As in the case of the empirical data set,
  the two-sided significance level $\alpha$ was set to~$0.05$.
  \new{Different methods and scenarios are compared based on observed confidence or credible interval coverage probabilities and lengths.}

\subsection*{Estimation in R}
  The software environment~\textsf{R} \citep{R} and two of its
  extensions, the \texttt{metafor} \citep{metafor2009,viechtbauer2010}
  and \texttt{bayesmeta} \citep{bayesmeta2015,Roever2017} packages are
  used with their default options. As no implementation in~\textsf{R}
  was found for the PL~estimation of Poisson-normal model we
  translated the steps described by B\"ohning \textit{et al.}
  \citep[][Ch. 2]{bohning2008meta} into \textsf{R}~code which is shown
  in the appendix.

%%%%%%%%%%%%%%%%
%% Results    %%
%%%%%%%%%%%%%%%%
\section*{Results}
\subsection*{Empirical Study} 
  Most (419; 80\%) of the 521~documents searched did not include a
  meta-analysis, because either the assignment was canceled (11; 2\%),
  the assignment had just started without results being available at
  the time of search (70; 13\%), no meta-analysis was included or
  accepted by the IQWiG (186; 36\%), no study (34; 7\%) or just one
  study (118; 23\%) was identified. Out of the remaining 102 documents
  which included at least one meta-analysis, 25 (5\%) did not include
  any binary meta-analysis, 19 (4\%) were network meta-analyses, and in
  18 (3\%) cases the first binary meta-analysis included at least one
  study with zero events. An overview over the identified 
% studies 
  \new{meta-analyses}
  is given in Table~\ref{tab:IQWiG_data}\new{; the data are also available online \citep{SeideEtAl2018}}.

  In the original publications, a slight majority of studies (26 of
  40) was analyzed using~RR as the effect measure.
  In the extracted data set, 21 out of the 40 meta-analyses (53\%)
  included only 2~studies, while 10 (25\%) consisted of three
  studies. Even in this small example, a common occurrence of 2- and
  3-study meta-analyses is found, which is also observed empirically
  by \citep{turner2012} and \citep{kontopantelis2013}. The
  distribution of study sizes and endpoints is also illustrated on the
  left panel in Figure~\ref{fig:descriptiveIQWiG}. With only two
  studies included, three methods coincide: the DL, the REML and the
  EB estimation \citep{rukhin2012}. As this is the case for a major
  share of the data set, these three methods are expected to show
  similar results in the analysis. The maximum number of studies
  observed is~18. The original analyses were based on the NNHM, and,
  with only the exceptions of the publications \mbox{A15-45},
  \mbox{S11-01} and \mbox{A11-30} performed using DL variance
  estimation.

  Imbalance in study sizes may influence the estimation of an overall
  treatment effect \citep{InthoutIoannidisBorm2014,RoeverKnappFriede2015,partlett2016}. As IntHout
  \textit{et al.}  \citep{InthoutIoannidisBorm2014} observe in an empirical study,
  such unequal study sizes are common in meta-analyses. In the data
  set extracted from IQWiG publications, ratios of sample sizes
  between the largest and the smallest study in a meta-analysis ranged
  from~1.0 up to~15.8, with a mean of~3.4 and a median of~1.9. Nearly
  half of the meta-analyses included at least one study twice as large
  as the smallest study.
  In the NNHM, study-specific variances~$\sigma_i^2$ should roughly be
  inversely proportional to sample sizes; imbalances in sample size
  then affect analysis via an imbalance in the~$\sigma_i$.
  Ratios of largest to smallest study sizes and variances using
  both effect measures for all studies are shown on the right panel in
  Figure~\ref{fig:descriptiveIQWiG}, where the ratio between the largest and
  the smallest value is ordered by the ratio of sample sizes in
  descending order. It can be observed that the ratios of the
  variances of ORs seem to vary more when study sizes are unbalanced
  than those of the~RRs. However, they both roughly follow the same
  pattern as the ratio of study sizes.

  The extracted data set is then analyzed based on the models and
  methods described above. As \mbox{$2\!\times\!2$ tables} are
  available for all studies, both effect measures are used to
  summarize the individual meta-analyses and to evaluate the influence
  of the choice of effect measure on the estimation results. The
  ratios of point estimates of the different methods against the
  standard DL approach are illustrated by the first row of
  Figure~\ref{fig:ResultsEmpirical} where the RR is displayed in the
  left and the OR in the right panel. As expected, DL, REML and EB
  estimation coincide in the majority of cases including only two
  studies \citep{rukhin2012}. These three estimators are also observed
  to behave comparable when more than two studies are included, as do
  the point estimates of the Bayesian approach. The greatest deviation
  from the standard DL approach is observed in the \new{GLMMs} 
  in both effect measures. In the case of OR as an effect
  measure, UM.FS and UM.RS perform comparable. CM.EL~estimation does
  not converge in all cases, however, in the cases where convergence
  was achieved, it is in line with DL estimation. The CM.AL however,
  is in general different from the DL estimation. The Poisson-based
  results also differ considerably from the DL estimates.

  The length of confidence intervals for the frequentist and credible
  intervals in the Bayesian estimation are also of importance as it
  might not be possible to detect significant treatment effects if
  intervals are inconclusively wide. For both effect measures, all
  intervals and the discussed adjustments are shown in the second row
  in Figure~\ref{fig:ResultsEmpirical}. Again, the RR is displayed in
  the left and the OR in the right panel. The Bayesian credible
  intervals are generally wider than the unadjusted confidence
  intervals 
  and more similar to the adjusted ones with respect to the median
  length, but exhibiting less variability.

\subsection*{Simulation Study}
  Figures~\ref{fig:ResultsSimulationRR} and~\ref{fig:ResultsSimulationOR} 
  illustrate the coverage rates
  (first row) and lengths (second row) of the 95\% confidence or
  credible intervals of the different methods
  for the relative risks and odds ratios, respectively. 
  All
  results shown here exemplarily refer to the combination of 100
  participants per arm and study and a baseline event rate of 0.7. 
  \new{Results of the other scenarios may be found in the supplement.}
  The
  different methods are indicated by colours, while the different
  adjustments are indicated by the line type.

  \new{Non-convergence rates averaged over all scenarios and both effect measures are mostly negligible in the methods based on the normal likelihood on the log-scale (ML:~0.049\%, EB:~0.032\%, HN(1.0):~0.002\%, HN(0.5):~0.036\%). Estimation based on REML or the methods taking the distributional assumptions on the trial-arms lead to slightly higher non-convergence rates (REML:~0.43\%, UM.FS:~0.43\%, UM.RS:~0.22\%, CM.AL:~0.47\%). The only method with high non-convergence rates is CM.EL, with an average of~18\%.}
  None of the methods fully dominates the others over the range of the
  investigated scenarios. Estimation using CM.EL for the
  binomial-normal model however was computationally expensive and
  convergence was problematic in a large proportion of scenarios
  (using the default values), as has been noted before
  \citep{viechtbauer2005,JacksonEtAl2018}; these results are omitted
  here. 
  Coverage rates of all methods are comparable when either the
  number of studies included in each meta-analysis is sufficiently
  large or when the heterogeneity is absent or low
  (\mbox{$I^2\!\leq\!0.25$}). However, given the frequency with which
  2- or 3-study meta-analyses occur empirically in our example data
  set and others \citep{turner2012,kontopantelis2013} and the
  difficulties in the determination of the absence of heterogeneity
  \citep{RoeverKnappFriede2015} this is hardly relevant in practice. In general
  when study sizes were not balanced, coverage rates for all methods
  were substantially lower even in the presence of only low
  heterogeneity \new{in the simulation of OR, while Bayesian estimation and the adjustment of frequentist confidence intervals resulted in better coverage when RR was used. This might be due to the $I^2$~values translating to lower values of absolute heterogeneity in the latter case}. In the presence of heterogeneity, coverage could drop
  as low as 40\% for some extreme scenarios in both, frequentist and
  Bayesian estimation, resulting in high false-positive rates. 
%  In general, it was also observed that one large study led to lower coverage than one small study per meta-analysis in the frequentist methods, which is in line with \citep{InthoutIoannidisBorm2014}. 
  \new{In general, it was also observed that one large study tended to lead to lower coverage than one small study per meta-analysis in the frequentist methods when more than $k\!=\!2$ studies are present, which is in line with \citep{InthoutIoannidisBorm2014}.
  This effect is more noticeable in the unadjusted methods, in scenarios where the number of patients per arm ($n_i$) is small, or when heterogeneity ($I^2$) is large. When considering $k\!=\!2$~trials per meta-analysis, coverages are comparable (OR) or the effect is even reversed (RR).} 
  The frequentist
  methods based on the normal-normal hierarchical model perform
  similarly, or, in the case of two studies per meta-analysis, even
  identically \citep{rukhin2012}. In the case of heterogeneous data, in
  particular regarding small study sizes, all frequentist methods
  perform below the nominal coverage probability when confidence
  intervals are not adjusted. In the scenarios with unbalanced study
  sizes this is even more pronounced than in the balanced scenarios;
  this is in line with the findings of \citep{InthoutIoannidisBorm2014} and
  \citep{RoeverKnappFriede2015}. Coverage can, at the cost of interval width, be
  increased by either using the HKSJ or the mHKSJ adjustment, but the
  HKSJ adjustment yields in some scenarios coverage probabilities
  which are still below the nominal level \citep{InthoutIoannidisBorm2014}.

%Length of confidence or credible intervals:
  The length of confidence and credible intervals is illustrated in
  the second row of Figures~\ref{fig:ResultsSimulationRR} and~\ref{fig:ResultsSimulationOR}. 
  Bayesian
  credible intervals are, as in the case for the empirical data
  set, in general wider than the unadjusted confidence intervals from
  the frequentist estimations. 
%When compared to the adjusted confidence intervals, Bayesian credible intervals are, especially in the presence of heterogeneity and with only two studies per meta-analysis narrower. 
  \new{When compared to  adjusted confidence intervals, Bayesian credible intervals tend to be, especially in the presence of heterogeneity and with only two studies per meta-analysis, narrower than the frequentist intervals based on the normal-normal model as long as the number of patients per trial arm is small. However, when $n_i$ increases, this is no longer true for RR (in contrast to the scenarios using OR). These differences may be due to the fact that idential $I^2$~settings can imply very different (and sometimes possibly unrealistically large) magnitudes of heterogeneity values on the $\tau$~scale, as can also be seen in Table~\ref{tab:Heterogeneity}.}
  In these extreme scenarios, adjusted
  frequentist confidence intervals are observed to be inconclusively
  wide\new{, with the exception of BN-UM.FS and BN-UM.RS, and especially when estimation is based on the NNHM}\@. In the other scenarios, Bayesian credible and adjusted
  confidence intervals are comparable.

%% Mean bias:
%  Table~\ref{tab:bias} illustrates the average, minimum and maximum
%  bias of the mean effect estimate for every method and both effect
%  measures over the different scenarios. Empirically, the estimates
%  from CM.AL and PL were observed to differ most from the standard DL
%  estimation. When combining all numbers of studies and all levels of
%  heterogeneity for the presented scenario, these two methods show a
%  low mean bias that is comparable to that of the other methods. The
%  bias is negative for PL and positive for CM.AL\@.

%%%%%%%%%%%%%%%%
%% Discussion %%
%%%%%%%%%%%%%%%%
\section*{Discussion}
  In our empirical study we found that the majority of the
  40~meta-analyses extracted from publications of IQWiG included only
  two studies. This is in agreement with a much larger empirical
  investigation based on the Cochrane Library by Turner
  \textit{et~al.}  \citep{TurnerEtAl2012}. This finding emphasizes the
  need for methods appropriate for meta-analysis with 
% very 
  few
  studies. Furthermore, varying methods and / or effect measures lead
  to differences in the results for the 40~meta-analyses
  considered. This demonstrates that prespecification of methods as
  well as effect measures is important for controlling operating
  characteristics.
  \new{The problems encountered in meta-analyses of few studies may mostly be attributed to the estimation of heterogeneity, and in particular to the proper accounting for its uncertainty in constructing intervals for the combined effect. The difference in performance between different heterogeneity estimators is relatively small compared to the difference in whether or how heterogeneity uncertainty is propagated through to the effect estimate \citep{FriedeRoeverWandelNeuenschwander2017a}.}

  In the simulation study, coverage probability was below the nominal
  level for all frequentist methods in the presence of heterogeneity
  and few studies. This phenomenon is even more pronounced when
  studies included in a meta-analysis are of unequal size. However,
  coverage probabilities generally improve when confidence intervals
  are adjusted based on the Student-$t$-distribution. 
%  Bayesian methods result in better coverage in all scenarios. 
  \new{Bayesian methods mostly result in better coverage across all scenarios, except for some cases of very large heterogeneity (in terms of~$\tau$) where the coverage is slightly lower.} 
  Credible intervals are
  empirically and in the simulation study wider than unadjusted
  confidence intervals, 
% but narrower than adjusted ones
  \new{but considerably narrower than adjusted ones, with some exceptions when considering RRs and large numbers of patients per trial-arm}. Previous
  simulation studies comparing a more restricted set of methods
  including standard frequentist and Bayesian approaches only led to
  similar conclusions. The simulations presented here considering a
  wider set of methods show 
  that the issues entailed by the increased complexity of some likelihood-based approaches may often outweigh their expected advantages \citep{BenderEtAl2018}.
  However, confidence
  intervals based on the \new{GLMMs} for example are in
  general slightly narrower than those from other frequentist
  methods. Furthermore, certain maximum-likelihood methods turned out to suffer from
  frequent numerical problems in the setting with few studies. To our
  knowledge, this has not been described previously.

%Limitations
  Our empirical investigation did not consider all IQWiG reports, but
  only the most recent 40~meta-analyses at the time of extraction. A
  consideration of all meta-analyses might have led to a more complete
  picture, but was not feasible with the resources of this project as
  no specific funding was available. Furthermore, the simulation study
  could have been enriched by additional methods. For instance, we
  only considered Bayesian two-stage approaches but did not include
  Bayesian approaches utilizing the full information of the $2\!\times\!2$
  tables. 
  The latter was considered recently by \citep{GuenhanEtAl2018} in the
  context of network meta-analyses, where pairwise meta-analysis would
  be a special case.
  As for the likelihood methods, we would expect that the results of
  the one-stage approach are overall quite similar to those of the
  two-stage approach considered, maybe with the potential of some
  small improvements.
  \new{As discussed in the context of the simulation setup, a pre-specified $I^2$~value may correspond to rather different $\tau$~values, depending on the circumstances (see also Table~\ref{tab:Heterogeneity}). Consequently, one may generally expect larger $I^2$~values for log-RR endpoints, and smaller $I^2$~values for log-OR endpoints, while heterogeneity priors are probably best discussed at the scale of $\tau$~values (a prior specification in terms of $I^2$ would be possible \citep{Roever2017}, but this would be hard to motivate). By relating the heterogeneity to the $\tau$~value, the question to consider is \emph{by what factor} the true RRs or ORs $\theta_i$ are expected to differ solely due to between-trial heterogeneity \citep[Sec.~5.7.3]{spiegelhalter2004}, and the reasonably expected range should then be covered by the prior. For example, a heterogeneity of $\tau\!=\!1.0$ implies that the central 95\% of true study means ($\theta_i$) span a range of a factor of~50 \citep{spiegelhalter2004,FriedeRoeverWandelNeuenschwander2017a}. The HN(0.5)-prior confines~$\tau$ to values below~1.0 with roughly 95\% probability, while the HN(1.0)-prior constitutes a conservative variation that instead allows for twice as large heterogeneity, implying a plausible range of roughly up to factor of $50^2=2500$.}

  \new{The limits of applicability of approximate meta-analysis methods have been discussed from the perspective of the NNHM by Jackson and White \citep{JacksonWhite2018}. In the limit of many studies (large~$k$) and large sample sizes (large~$n_i$), the normal approximation usually works well. It starts breaking down, however, when the number of studies ($k$) gets too small. The problem then is related to the estimation of heterogeneity ($\tau$) and proper accounting for the associated uncertainty; inference would still be exact if the heterogeneity was known.
In the frequentist context, use of the HKSJ adjustment helps, especially if the study-specific standard errors are roughly balanced \citep{VeronikiEtAl2018}.
This is not so much of a problem when Bayesian methods along with reasonable priors are used; these methods yield valid inference irrespective of the number of included studies \citep{RoeverFriede2018}.
Problems also arise when events are rare or sample sizes ($n_i$) are small. In either case, the chances of observing few or no events in a treatment group increase, and normal approximations to the likelihood break down. In such situations, a solution might be to resort to exact likelihoods respecting the discrete nature of the data, for example a GLMM, which may again be done in frequentist or Bayesian frameworks \citep{bohning2008meta,viechtbauer2010,GunhanRoeverFriede2018}.}

%%%%%%%%%%%%%%%%%
%% Conclusions %%
%%%%%%%%%%%%%%%%%
\section*{Conclusions}

In the presence of between-study heterogeneity, especially with unbalanced study sizes, caution is needed in applying meta-analytical methods to 
%(very) 
few studies, as either coverage probabilities of intervals may be compromised, or they may be inconclusively wide. Bayesian estimation with sensibly chosen prior for the between-study heterogeneity may offer a compromise and promising alternative.

%%%%%%%%%%%%%%%%%%%%%%%%%%%%%%%%%%%%%%%%%%%%%%
%%                                          %%
%% Backmatter begins here                   %%
%%                                          %%
%%%%%%%%%%%%%%%%%%%%%%%%%%%%%%%%%%%%%%%%%%%%%%

\begin{backmatter}

\section*{Abbreviations}
  For abbreviations of model variations used, see also Table~\ref{tab:ModelAbbrev}.
  \begin{tabular}{ll}
    %\toprule
    DL    & DerSimonian-Laird\\
    EB    & empirical Bayes\\
    GLMM  & generalized linear mixed model\\
    HKSJ  & Hartung-Knapp-Sidik-Jonkman\\
    IPD   & individual patient data\\
    IQWiG & Institut f\"{u}r Qualit\"{a}t und Wirtschaftlichkeit im Gesundheitswesen\\
          & (Institute for Quality and Efficiency in Health Care)\\
    mHKSJ & modified HKSJ\\
    ML    & maximum likelihood\\
    NNHM  & normal-normal hierarchical model\\
    OR    & odds ratio\\
    PL    & profile likelihood\\
    REML  & restricted ML\\
    RR    & risk ratio\\
    %\bottomrule
  \end{tabular}

\section*{Ethics approval and consent to participate}
  Not applicable.

\section*{Consent for publication}
  Not applicable.

\section*{Availability of data and material}
  The IQWiG dataset used in current study is 
  \new{available online \citep{SeideEtAl2018}}.

\section*{Competing interests}
  The authors declare that they have no competing interests.

\section*{Funding}
  Not applicable.

\section*{Author's contributions}
  TF conceived the concept of this study. SES carried out the
  simulations and drafted the manuscricpt. CR critically reviewed and
  made substantial contributions to the manuscript. All authors
  commented on and approved the final manuscript.

\section*{Acknowledgements}
  The authors would like to thank Ralf Bender (IQWiG, K\"{o}ln,
  Germany) for helpful comments on the manuscript.

%%%%%%%%%%%%%%%%%%%%%%%%%%%%%%%%%%%%%%%%%%%%%%%%%%%%%%%%%%%%%
%%                  The Bibliography                       %%
%%                                                         %%
%%  Bmc_mathpys.bst  will be used to                       %%
%%  create a .BBL file for submission.                     %%
%%  After submission of the .TEX file,                     %%
%%  you will be prompted to submit your .BBL file.         %%
%%                                                         %%
%%                                                         %%
%%  Note that the displayed Bibliography will not          %%
%%  necessarily be rendered by Latex exactly as specified  %%
%%  in the online Instructions for Authors.                %%
%%                                                         %%
%%%%%%%%%%%%%%%%%%%%%%%%%%%%%%%%%%%%%%%%%%%%%%%%%%%%%%%%%%%%%

% if your bibliography is in bibtex format, use those commands:
\bibliographystyle{bmc-mathphys} % Style BST file (bmc-mathphys, vancouver, spbasic).
\bibliography{LikelihoodMA-Refs}      % Bibliography file (usually '*.bib' )
% for author-year bibliography (bmc-mathphys or spbasic)
% a) write to bib file (bmc-mathphys only)
% @settings{label, options="nameyear"}
% b) uncomment next line
%\nocite{label}

% or include bibliography directly:
% \begin{thebibliography}
% \bibitem{b1}
% \end{thebibliography}

%%%%%%%%%%%%%%%%%%%%%%%%%%%%%%%%%%%
%%                               %%
%% Figures                       %%
%%                               %%
%% NB: this is for captions and  %%
%% Titles. All graphics must be  %%
%% submitted separately and NOT  %%
%% included in the Tex document  %%
%%                               %%
%%%%%%%%%%%%%%%%%%%%%%%%%%%%%%%%%%%

%%
%% Do not use \listoffigures as most will included as separate files

\onecolumn

 \section*{Figures}
%\section*{Figure legends}
  
  \begin{figure}[h!]
    \includegraphics[width = 0.99\textwidth]{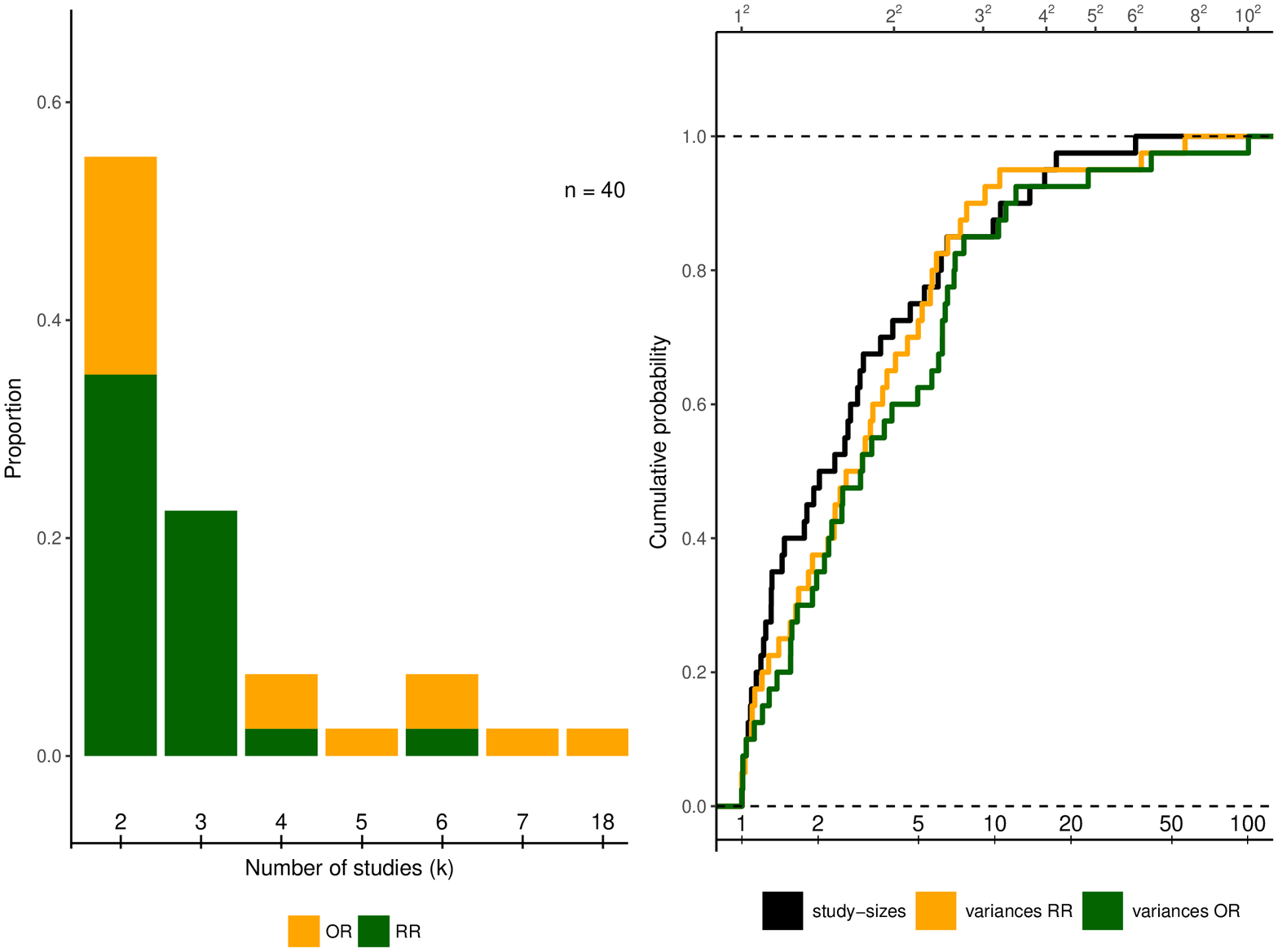}
    \caption{\label{fig:descriptiveIQWiG}\csentence{Characteristics of the data set extracted from IQWiG publications.} Left side: Proportions of number of studies included per meta-analysis out of n = 40. Colours indicate the effect measure used in the original publications. Right side: Empirical distribution function for the proportion of study sizes (largest vs. smallest per meta-analysis, black) and the proportion of study-specific variances (largest vs. smallest per meta-analysis) for the log-transformed~RR (green) and the log-transformed~OR (orange). All meta-analyses are included for both effect measures.}
\end{figure}

  \begin{figure}[h!]
    \includegraphics[width = 0.98\textwidth]{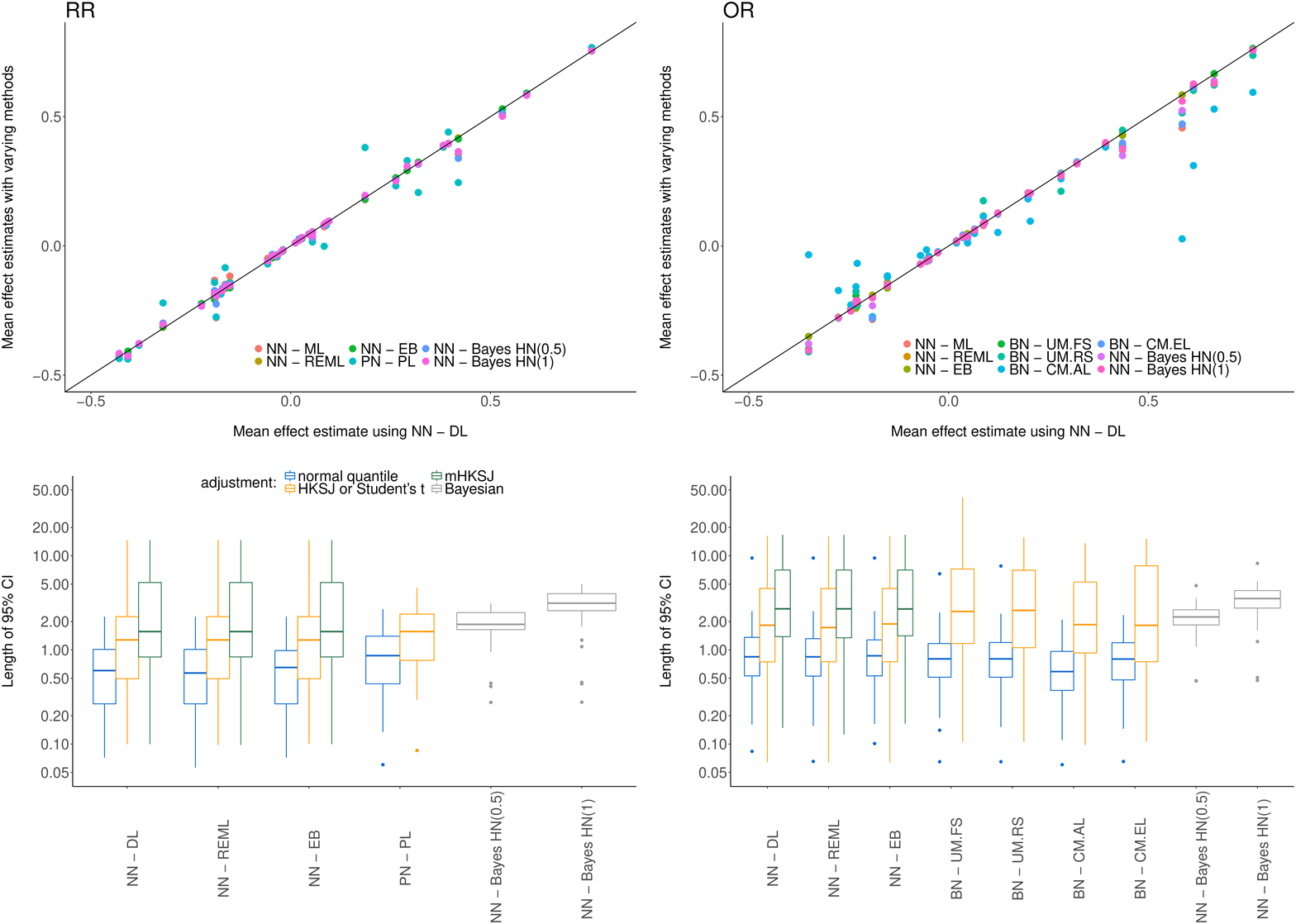}
    \caption{\label{fig:ResultsEmpirical}\csentence{Estimates of the combined treatment effect and lengths of confidence or credible intervals for both effect measures, empirical data set.} The first row shows the treatment effect estimates for the RR (left column) and the OR (right column) compared to the standard DL approach. Colours indicate the various methods. The second row illustrates the length of confidence or credible intervals and the respective adjustments, again for the RR (left column) and the OR (right column).}
  \end{figure}
  
  \begin{figure}[h!]
    \includegraphics[width = 0.72\textwidth]{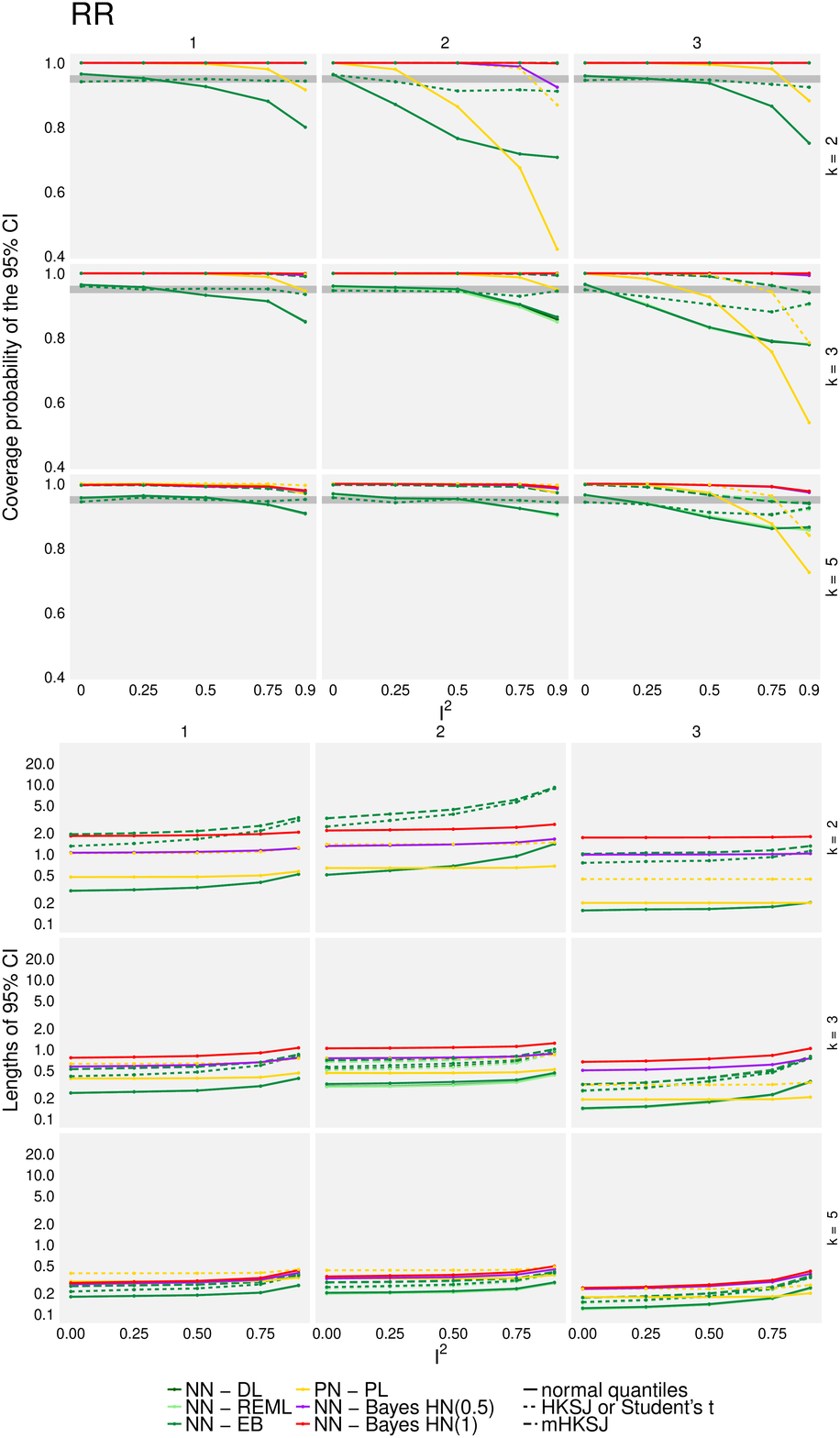}
    \caption{\label{fig:ResultsSimulationRR}\csentence{Coverage probabilities and lengths of 95\% confidence or credible intervals for the overall effect for RR~effects based on the simulated data.} The top panel shows the coverage probabilities of treatment effect CIs for the different methods (colours) and adjustments (line types). \new{The grey area indicates the range expected with 95\% probability if the coverage is accurate.} The bottom panel similarly shows the lengths of 95\% confidence or credible intervals. Results are illustrated for a study size of $n_i\!=\!100$ and a baseline event probability $p_0\!=\!0.7$, and are based on 2000~replications per scenario. CM.EL is omitted due to low convergence rates.}
  \end{figure}

  \begin{figure}[h!]
    \includegraphics[width = 0.72\textwidth]{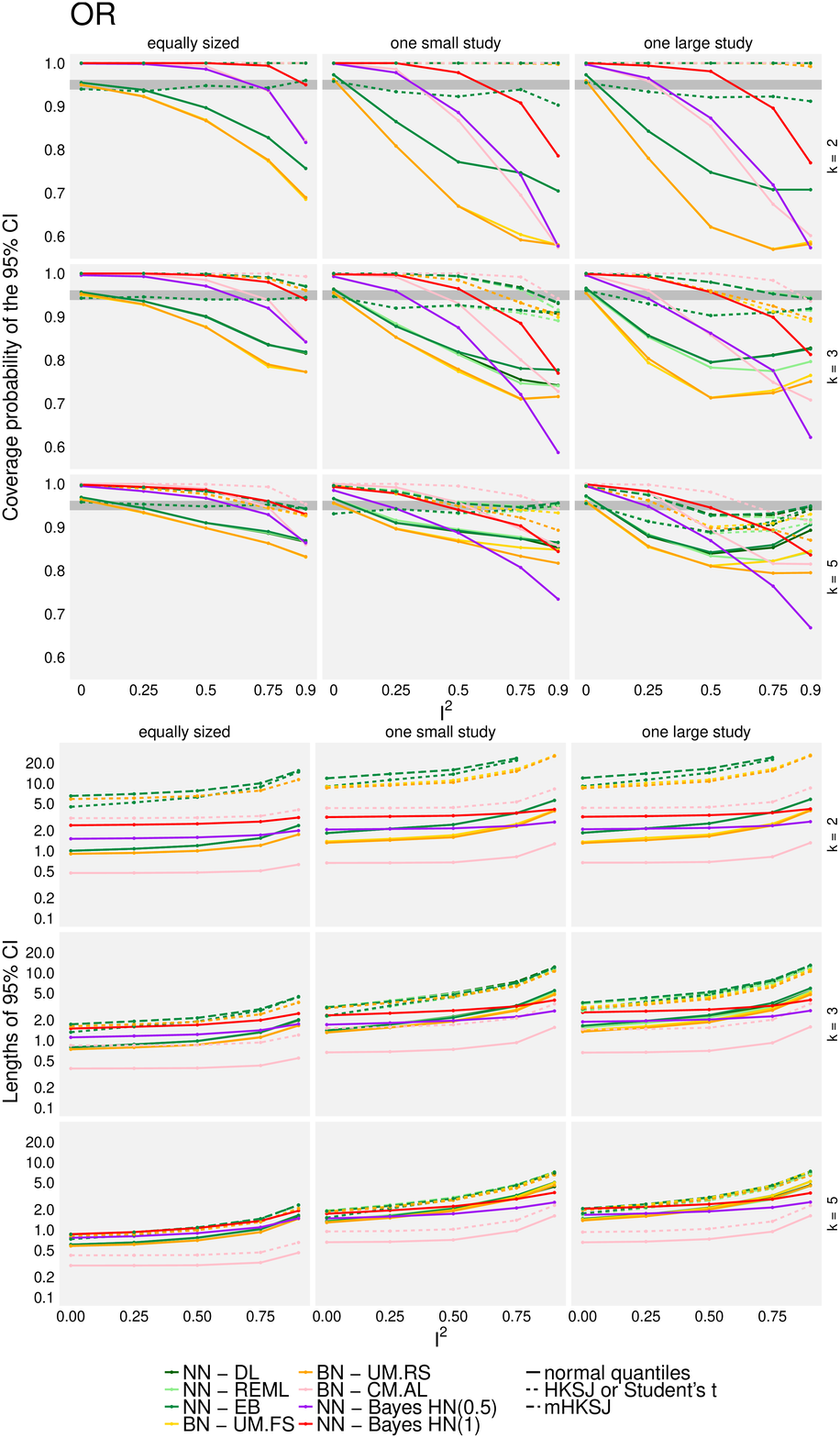}
    \caption{\label{fig:ResultsSimulationOR}\csentence{Coverage probabilities and lengths of 95\% confidence or credible intervals for the overall effect for OR~effects based on the simulated data.} The top panel shows the coverage probabilities of treatment effect CIs for the different methods (colours) and adjustments (line types). \new{The grey area indicates the range expected with 95\% probability if the coverage is accurate.} The bottom panel similarly shows the lengths of 95\% confidence or credible intervals. Results are illustrated for a study size of $n_i\!=\!100$ and a baseline event probability $p_0\!=\!0.7$, and are based on 2000~replications per scenario. CM.EL is omitted due to low convergence rates. }
  \end{figure}

  \twocolumn

%%%%%%%%%%%%%%%%%%%%%%%%%%%%%%%%%%%
%%                               %%
%% Tables                        %%
%%                               %%
%%%%%%%%%%%%%%%%%%%%%%%%%%%%%%%%%%%

%% Use of \listoftables is discouraged.
%%
\onecolumn
\section*{Tables}
\begin{table}[!h]
\centering
\caption{\label{tab:IQWiG_data}Data extracted from IQWiG publications \new{\citep{SeideEtAl2018}}.}
\begin{tabular}{llllrccc}
  \toprule
        &            &      &          &      & \textbf{Number of}     & \textbf{Effect}\\ 
    \textbf{No.} & \textbf{Identifier} & \textbf{Date} & \textbf{Endpoint} & \textbf{Page} & \textbf{studies ($k$)} & \textbf{measure}\\ 
  \midrule
  \phantom{0}1 & N15-06 & 2017-03 & morning pain & 85 & 5 & OR  \\
  \phantom{0}2 & N15-11 & 2017-03 & ear infection & 62 & 2 & OR \\
  \phantom{0}3 & S15-02 & 2017-01 & mortality & 53 &   2 & OR \\
  \phantom{0}4 & D15-02 & 2017-01 & mortality & 74 & 2 & OR \\
  \phantom{0}5 & A16-71 & 2016-12 & morbidity & 5 & 6 & OR \\ \addlinespace[1.0ex]
  \phantom{0}6 & A16-38 & 2016-12 & vomiting & 4 & 2 & RR \\
  \phantom{0}7 & P14-03 & 2016-11 & breast cancer screening & 55 & 3 & RR \\
  \phantom{0}8 & N14-02 & 2016-08 & remission from anxiety disorder & 127 &   2 & OR \\
   \phantom{0}9 & A16-30 & 2016-08 & AIDS-defining events & 103 &   2 & RR \\
            10 & N15-07 & 2016-08 & ejaculation dysfunction & 89 &   4 & OR \\ \addlinespace[1.0ex]
            11 & A16-11 & 2016-06 & serious adverse events & 86 &   2 & RR \\
            12 & A10-03 & 2016-04 & serious adverse events & 89 &   2 & OR \\
            13 & A15-57 & 2016-02 & St.\ George's Respiratory Questionnaire response & 22 &   2 & RR  \\
            14 & A15-45 & 2016-01 & morbidity & 24 &   2 & OR \\
            15 & A15-31 & 2015-11 & mortality & 87 &   2 & RR  \\ \addlinespace[1.0ex]
            16 & A15-25 & 2015-10 & serious adverse events & 89 &   2 & RR \\
            17 & A15-21 & 2015-07 & mortality & 16 &   2 & RR \\
            18 & S13-04 & 2015-05 & screening for abdominal aortic aneurysm & 71 &   4 & OR \\
            19 & A15-06 & 2015-05 & morbidity & 96 &   3 & RR \\
            20 & A15-05 & 2015-03 & morbidity & 4 &   2 & RR \\ \addlinespace[1.0ex]
            21 & A14-38 & 2015-01 & serious adverse events & 65 &   3 & RR \\
            22 & A14-25 & 2014-11 & serious adverse events & 115 &   2 & RR \\
            23 & A14-22 & 2014-10 & Transition Dyspnea Index responder & 67 &   2 & RR \\
            24 & A14-19 & 2014-09 & urge to urinate & 75 &   3 & RR \\
            25 & A14-18 & 2014-09 & persistent virological response (SVR24) & 194 &   3 & RR \\ \addlinespace[1.0ex]
            26 & S13-03 & 2014-06 & participants with cervical intraepithelial neoplasia 3+ & 15 &   6 & RR \\
            27 & A13-29 & 2013-10 & metformidosis & 15  &   3 & RR \\
            28 & A10-01 & 2013-08 & remissions & 1183 &   2 &   OR \\
            29 & A13-20 & 2013-08 & visual acuity & 28 &   3 & RR \\
            30 & S11-01 & 2013-07 & bowel cancer & 61 &   7 & OR \\ \addlinespace[1.0ex]
            31 & A13-23 & 2013-06 & mortality & 15 &   2 & RR \\
            32 & A13-05 & 2013-04 & full recovery & 19 &   4 & RR \\
            33 & A05-10 & 2013-04 & cardiovascular death & 75 &   3 & RR \\
            34 & A12-19 & 2013-03 & ocular adverse event & 17 & 2 & RR \\
            35 & A05-18 & 2012-08 & serious adverse events & 67 & 18 & OR \\ \addlinespace[1.0ex]
            36 & A12-10 & 2012-07 & adverse events & 20 & 3 & RR \\
            37 & A12-03 & 2012-04 & loss of transplant & 23 & 2 & RR \\
            38 & A12-04 & 2012-04 & virus occurrence & 22 & 3 & RR \\
            39 & A09-05 & 2012-04 & Alzheimer’s Disease Assessment Scale & 51 & 6 & OR \\ 
            40 & A11-30 & 2012-03 & mortality & 24 & 2 & OR \\
  \bottomrule
\end{tabular}
\end{table}

%\twocolumn

\begin{table}[!ht]
  \caption{\label{tab:ModelAbbrev}Abbreviations used for analysis models.}  
  \centering 
  \begin{tabular}{ll}
  \toprule
  NN-DL    & Normal-normal (NN) model using the DerSimonian-Laird (DL) heterogeneinty estimator \\
  NN-REML  & NN model using the restricted maximum-likelihood (REML) estimator \\
  NN-EB    & NN model using the empirical-Bayes (EB) estimator \\
  PN-PL    & Poisson model using profile likelihood (PL) estimation \\
  BN-UM.FS & Binomial model using unconditional logistic regression and fixed study (nuisance) parameters \\
  BN-UM.RS & Binomial model using unconditional logistic regression and random study (nuisance) parameters \\
  BN-CM.EL & Conditional (hypergeometric) model (exact likelihood)\\
  BN-CM.AL & Conditional (hypergeometric) model (approximate likelihood)\\
  NN-Bayes HN(0.5) & NN Bayesian model using a half-normal heterogeneity prior with scale 0.5 \\
  NN-Bayes HN(1.0) & NN Bayesian model using a half-normal heterogeneity prior with scale 1.0 \\
  \bottomrule
  \end{tabular}
\end{table}

\begin{table}[!ht]
  \caption{\label{tab:SimulationParameter}Parameters of the simulation for both effect measures, i.e., relative risk and odds ratio.}
  \centering
  \begin{tabular}{ll}
  \toprule
  \textbf{Parameter} & \textbf{Values} \\
  \midrule
  Effect measure ($\theta_i$)    &  RR, OR\\
  Design                         &  \makecell[tl]{equally sized studies, \\
                                                  one small study ($\frac{1}{10}$ size) \\
                                                  one large study ($10$-fold size)}\\
  Observations per study arm ($n_i$)  &  25, 50, 100, 250, 500, 1000\\
  Number of studies ($k$)        &  2, 3, 5, 10 \\
%  Replications                   &  2000 \\
  Event rates ($p_0$)            &  0.1, 0.3, 0.5, 0.7, 0.9 \\
  Level of heterogeneity ($I^2$) & \makecell[tl]{no heterogeneity: 0.00 \\
		                                 low heterogeneity: 0.25 \\
                                                 moderate heterogeneity: 0.50\\
                                                 high heterogeneity: 0.75 \\
                                                 very high heterogeneity: 0.90}\\
  \bottomrule
  \end{tabular}
\end{table}

\begin{table}[!ht]
  \caption{\label{tab:Heterogeneity}Absolute heterogeneity values
    ($\tau$) corresponding to relative settings ($I^2$) used in the
    simulations that are shown in Figures~\ref{fig:ResultsSimulationRR} and~\ref{fig:ResultsSimulationOR}.}  
  \centering
  \begin{tabular}{lccccccc}
  \toprule
  && \multicolumn{3}{c}{\textbf{relative risk (RR)}} 
    & \multicolumn{3}{c}{\textbf{odds ratio (OR)}} \\
  \cmidrule(lr){3-5}  \cmidrule(lr){6-8}
  & $I^2$ & equal & one small & one large & equal & one small & one large \\
  \midrule
  $k\!=\!2$ & 0.25 &  0.0534 & 0.1254 & 0.0396 &  0.1781 & 0.4179 & 0.1321 \\
            & 0.50 &  0.0926 & 0.2171 & 0.0687 &  0.3086 & 0.7237 & 0.2289 \\
            & 0.75 &  0.1604 & 0.3761 & 0.1189 &  0.5345 & 1.2536 & 0.3964 \\
            & 0.90 &  0.2777 & 0.6514 & 0.2060 &  0.9258 & 2.1712 & 0.6866 \\[1ex]
  $k\!=\!3$ & 0.25 &  0.0534 & 0.1069 & 0.0447 &  0.1781 & 0.3563 & 0.1491 \\
            & 0.50 &  0.0926 & 0.1852 & 0.0775 &  0.3086 & 0.6172 & 0.2582 \\
            & 0.75 &  0.1604 & 0.3207 & 0.1342 &  0.5345 & 1.0690 & 0.4472 \\
            & 0.90 &  0.2777 & 0.5549 & 0.2324 &  0.9258 & 1.8516 & 0.7746 \\[1ex]
  $k\!=\!5$ & 0.25 &  0.0534 & 0.0844 & 0.0484 &  0.1781 & 0.2981 & 0.1613 \\
            & 0.50 &  0.0926 & 0.1549 & 0.0838 &  0.3086 & 0.5164 & 0.2795 \\
            & 0.75 &  0.1604 & 0.2683 & 0.1452 &  0.5345 & 0.8944 & 0.4840 \\
            & 0.90 &  0.2777 & 0.4648 & 0.2515 &  0.9258 & 1.5491 & 0.8384 \\
  \bottomrule
  \end{tabular}
\end{table}

%\begin{table}[!ht]
%\caption{\label{tab:bias}Means and ranges of bias of treatment effect
%  estimates~$\hat{\theta}$ across simulation scenarios.}  
%\centering
%\begin{tabular}{lcc}
%  \toprule
%  \textbf{Method} & \textbf{relative risk (RR)} & \textbf{odds ratio (OR)} \\
%  \midrule
%   NN-DL            & $\phantom{-}0.0027$ ($-1.1353$; $0.9426$) & $\phantom{-}0.0045$ ($-3.6621$; $3.3251$)\\
%   NN-REML          & $\phantom{-}0.0025$ ($-1.1354$; $0.9428$) & $\phantom{-}0.0059$ ($-3.6623$; $3.3252$) \\
%   NN-EB            & $\phantom{-}0.0027$ ($-1.1353$; $0.9526$) & $\phantom{-}0.0059$ ($-3.6624$; $3.3252$) \\
%   PN-PL            &           $-0.0002$ ($-1.5150$; $1.5991$) & \\
%   BN-UM.FS         &                                           & $\phantom{-}0.0038$ ($-4.2890$; $3.0263$) \\
%   BN-UM.RS         &                                           & $\phantom{-}0.0008$ ($-3.2653$; $2.7860$)\\
%   BN-CM.AL         &                                           & $\phantom{-}0.0030$ ($-1.2249$; $1.3640$)\\
%   NN-Bayes HN(0.5) &           $-0.0043$ ($-0.2571$; $0.2249$) &           $-0.0025$ ($-2.1473$; $2.2211$) \\
%   NN-Bayes HN(1.0) &           $-0.0016$ ($-0.4435$; $0.6932$) &           $-0.0073$ ($-2.1770$; $2.0372$)\\
%  \bottomrule
%\end{tabular}
%\end{table}

\begin{table}
\caption{\label{tab:simulationprocedure}Generation of data sets in the simulation study.}
\center
\newgeometry{top=1.25in, left=0cm, bottom = 1.25in, right=4.5cm}
\resizebox{!}{\linewidth}{
\begin{tikzpicture}[fill=blue, ultra thick, transform shape]
\tikzstyle{myarrows}=[line width=1mm,draw=blue,-triangle 45,postaction={draw, line width=4mm, -}]

\linespread{1.4}

		\node (shape1) at (5,30) [draw] {\makecell{Setting of the parameters\\
																			 $I^2$, $k$, $n_i$, $P_0$, $case$}}; 
																			
		\node (shape1) at (5,27) [draw] {\makecell{Determining the vector of study sizes from:\\
																			 $n_i$ and $case$}}; 																	
																			
		\node (shape1) at (5,23) [draw]	{\makecell{Calculation of the variance $\tau^2 = \epsilon^2 \frac{I^2}{1-I^2}$ \\
																								 \textbf{{with RR:} $\epsilon^2 = \frac{1}{k} \sum{\frac{1}{n_i} (\frac{2}{P_0} - 2 )}$  \hspace{0.5cm}} \\
																								 \textbf{{with OR:} $\epsilon^2 = \frac{1}{k} \sum{\frac{1}{n_i} (\frac{2}{P_0} + \frac{2}{1 - P_0} )}$} \\
																									from $I^2$, $n_i$ and $k$, $P_0$}};
																									
		\node (shape1) at (5,19.5) [draw]	{Generation of the true trial effect size $\theta_i$ from a normal distribution ($\mathrm{N}(0, \tau^2)$)};
		
		\node (shape1) at (5, 15.5) [draw]	{\makecell{Calculation of the true event rates $P_a$ and $P_b$ with:\\
		                                               \textbf{{RR:} $\log(P_a) = \log(P_0) - \frac{\theta_i}{2}$}\\
																									    \hspace{1cm} $\log(P_b) = \log(P_0) + \frac{\theta_i}{2}$ \\
																									 \textbf{\hspace{0.6cm} {OR:} $\log(\frac{P_a}{1-P_a}) = \log(\frac{P_0}{1-P_0}) - \frac{\theta_i}{2}$}\\
																									    \hspace{1.6cm} $\log(\frac{P_b}{1-P_b}) = \log(\frac{P_0}{1-P_0}) + \frac{\theta_i}{2}$
		                                               }};
																																																				
		\node (shape1) at (5,11.5) [draw]	{Generation of the observed event rates $p_a$ ($p_b$) from a Bernoulli distribution with $P_a$ ($P_b$)};
		
		\node (shape1) at (5,8.5) [draw]	{\makecell{Calculation of the events $x_{it}$ and $x_{ic}$,  the $\hat{\theta}_i$, and $\hat{\sigma}^2_{i,\hat{\theta}_i}$ from\\
																				       $p_a$, $p_b$ and $n_i$}};
		\node (shape1) at (14,15.5) [draw]	{\makecell{repeat with:\\
																									$rep = 2000$}};
		
		\node (shape1) at (5,5.5) [draw]	{Arranging into one data set};
		\node (shape1) at (16,27) [draw] {\makecell{For all combinations\\ of parameter}};

		\draw[->] (5,29) -- (5,28); \draw[<-] (7.5,30) -- (16,30);
		\draw[->] (5,26) -- (5,24.7); \draw[<-] (8.7,23) -- (14,23);
		\draw[->] (5,21.3) -- (5,19.85);
		\draw[->] (5,19.18) -- (5,17.55);
		\draw[->] (5,13.45) -- (5,11.85); 
		\draw[->] (5,11.15) -- (5,9.5); \draw (10.4,8.5) -- (14,8.5);
		\draw[->] (5,7.5) -- (5,5.85); \draw (7.7,5.5) -- (16,5.5);
		
		\draw[->] (14,8.5) -- (14,14.5); \draw (14,16.5) -- (14, 23);
		\draw[->] (16,5.5) -- (16,26); \draw (16,28) -- (16, 30);
		
\end{tikzpicture}
}
\restoregeometry
\end{table}

%%%%%%%%%%%%%%%%%%%%%%%%%%%%%%%%%%%
%%                               %%
%% Additional Files              %%
%%                               %%
%%%%%%%%%%%%%%%%%%%%%%%%%%%%%%%%%%%

    { \onecolumn
\section*{Additional Files}
  \subsection*{\textsf{R} code for Poisson PL estimation}
    Shown below is the \textsf{R}~code implementing profile likelihood
    (PL) estimation for the Poisson model according to B\"{o}hning \textit{et al.}
    \citep[][Ch. 2]{bohning2008meta} (see also \emph{Methods}
    section).
\begin{verbatim}
### Point Estimate under homogeneity
Estimate_PL_Poisson_RR <- function(x.t, n.t, x.c, n.c,
                                   log.effect = TRUE,
                                   s.crit = 0.0000001,
                                   theta.start = 1)
{
  w_thetas <- 0
  theta_homog <- theta.start
  theta_homog_n <- 1
  crit <- 1
  iter <- 0
  while(abs(unlist(crit)) > s.crit) {
    w_thetas <- 1/(n.c + theta_homog * n.t)  # weight of thetas
    # theta under homogeneity:
    theta_homog_n <- sum((1/(n.c + w_thetas * n.t)) * x.t * n.c) /
                     sum((1/(n.c + w_thetas * n.t)) * x.c * n.t)
    crit <- theta_homog - theta_homog_n
    theta_homog <- theta_homog_n
    iter <- iter + 1
  }
  if (log.effect) {
    return(c(log(theta_homog), as.integer(iter)))
  } else {
    return(c((theta_homog), as.integer(iter)))
  }
}


### Variance of the point estimate under homogeneity
Variance_PL_Poisson_RR <- function(x.t, n.t, x.c, n.c,
                                   theta, log.effect=TRUE)
{
  if (! log.effect) {
    first_term  <- sum(x.t)/ (theta^2)
    second_term <- - sum(((x.c + x.t) * (n.t^2)) /((n.c + theta * n.t)^2))
    return((-(first_term + second_term))^-1)
  } else {
    theta <- exp(theta)
    alpha <- (theta * n.t)/(n.c + theta * n.t)
    x_all <- x.t + x.c
    return(sum((x_all * alpha * (1-alpha)))^-1)
  }
}


### Confidence Interval for the point estimate under homogeneity
CI_PL_Poisson_RR <- function(x.t, n.t, x.c, n.c, theta,
                             log.effect=TRUE, distr)
{
  if (distr == "normal") {
    quant <- qnorm(0.975)
  }
  if (distr == "t-distributed") {
    quant <- qt(0.975, length(x.t))
  }
  if (log.effect) {
    variance <- (Variance_PL_Poisson_RR(x.t,n.t,x.c,n.c,theta,log.effect=TRUE)
                 + tau2_theta_Poisson(x.t,n.t,x.c,n.c,theta,log.effect=TRUE))
    lower <- theta - quant * sqrt(variance)
    upper <- theta + quant * sqrt(variance)
    return(c(lower, upper))
  } else {
    theta <- log(theta)
    variance <- (Variance_PL_Poisson_RR(x.t,n.t,x.c,n.c,theta,log.effect=TRUE)
                 + tau2_theta_Poisson(x.t,n.t,x.c,n.c,theta,log.effect=TRUE))
    lower <- theta - quant * sqrt(variance)
    upper <- theta + quant * sqrt(variance)
    return(c(exp(lower), exp(upper)))
  }
}


### 3. Estimator for the heterogeneity tau^2
mu_poisson <- function(x.t, n.t, x.c, n.c, theta)
{
  if (any(x.t == 0) | any(x.t == n.t)) {
    x.t <- x.t + 0.5
    n.t <- n.t + 0.5
  }
  if (any(x.c == 0) | any(x.c == n.c)) {
    x.c <- x.c + 0.5
    n.c <- n.c + 0.5
  }
  y <- x.t
  N <- x.t + x.c
  return(sum(y)/sum(N))
}


### Variance of p
tau2_p_Poisson <- function(x.t, n.t, x.c, n.c, theta)
{
  y <- x.t
  N <- x.t + x.c
  p <- (theta*n.t)/(n.c + theta*n.t) 
  numerator <- (sum((y-N*mu_poisson(x.t, n.t, x.c, n.c, theta))^2)
                -(sum(N))*mu_poisson(x.t, n.t, x.c, n.c, theta)
                 *(1-mu_poisson(x.t, n.t, x.c, n.c, theta)))
  denominator <- sum(N*(N-1))
  return(numerator/denominator)
}


### Between-study variance
tau2_theta_Poisson <- function(x.t, n.t, x.c, n.c, theta, log.effect=TRUE)
{
  if (log.effect) {
    result <- (tau2_p_Poisson(x.t, n.t, x.c, n.c, theta)
               / (1-mu_poisson(x.t, n.t, x.c, n.c, theta))^4)
    if (any(result < 0)) result[result < 0] <- 0
  } else {
    result <- (tau2_p_Poisson(x.t, n.t, x.c, n.c, theta)
               / (((mu_poisson(x.t, n.t, x.c, n.c, theta)^2)
                   *((1 - mu_poisson(x.t, n.t, x.c, n.c, theta))^2))))
    if (any(result < 0)) result[result < 0] <- 0
  }
  return(result)
}
\end{verbatim}
      \twocolumn }

\end{backmatter}
\end{document}